 \documentclass[useAMS, ]{mn2e}
 \usepackage{graphicx,amsmath}

\begin{document}

\title[The investigation of absolute proper motions of the XPM Catalogue]
  {The investigation of absolute proper motions of the XPM Catalogue }
\author[P.N. Fedorov et al.]
  {P.N.~Fedorov,$^1$ V.S.~Akhmetov,$^1$
  \newauthor V.V.~Bobylev,$^2$ A.T.~Bajkova $^2$ \\
  $^1$Institute of Astronomy of Kharkiv National University,
  Sums'ka 35, 61022 Kharkiv, Ukraine \\
E-mail: pnf@astron.kharkov.ua, akhmetov@astron.kharkov.ua\\
  $^2$Central astronomical observatory at Pulkovo of RAS,
 Pulkovskoye chaussee 65/1, 196140, Saint-Petersburg, Russia.\\
 E-mail: vbobylev@gao.spb.ru, anisabajkova@rambler.ru}

 \date{Accepted 2010 March 10. Received 2010 March 10; in original form 2010 March 10}
 \pagerange{\pageref{firstpage}--\pageref{lastpage}} \pubyear{2010}
 \def\LaTeX{L\kern-.36em\raise.3ex\hbox{a}\kern-.15em
    T\kern-.1667em\lower.7ex\hbox{E}\kern-.125emX}
 \newtheorem{theorem}{Theorem}[section]
 \def\degr{^\circ}
 \label{firstpage}
 \maketitle

\begin{abstract}
The XPM-1.0 is the regular version of the XPM catalogue. In
comparison with XPM the astrometric catalogue of about 280
millions stars covering entire sky from $-90^\circ$ to $+90^\circ$
in declination and in the magnitude range $10^{m}<B<22^{m}$ is
something improved. The general procedure steps were followed as
for XPM, but some of them are now performed on a more
sophisticated level. The XPM-1.0 catalogue contains star
positions, proper motions, 2MASS and USNO photometry of about 280
millions of the sources. We present some investigations of the
absolute proper motions of XPM-1.0 catalogue and also the
important information for the users of the catalogue. Unlike
previous version, the XPM-1.0 contains the proper motions over the
whole sky without gaps. In the fields, which cover the zone of
avoidance or which contain less than of 25 galaxies a quasi
absolute calibration was performed. The proper motion errors are
varying from 3 to 10 mas/yr, depending on a specific field. The
zero-point of the absolute proper motion frame (the absolute
calibration) was specified with more than 1 million galaxies from
2MASS and USNO-A2.0. The mean formal error of absolute calibration
is less than 1 mas/yr.
\end{abstract}

\section{Introduction}

In this work we describe still some steps towards the main goal
--- to create the most comprehensive catalogue of absolute proper
motions of stars --- XPM (Fedorov, Myznikov \& Akhmetov, 2009,
hereafter Paper I), using the extragalactic reference frame
defined by the faint galaxies.

As is well known, there are few catalogues of the absolute proper
motions of stars, while there are no catalogues that would cover
the whole celestial sphere. The southern hemisphere is supplied
with the data especially poorly, since there is a single catalogue
of absolute proper motions for the region southward of
$-45^\circ$, SPM1 (Platais et al., 1998), which covers the area
approximately 720 square degrees near the South Pole. The limiting
apparent stellar magnitude does not exceed 18$^{m}$ in all the
catalogues. They are all based on photographic observations made
in the 20-th century. The most known of them are the GPM (Rybka \&
Yatsenko, 1997; I/285 CDS), the PUL2 (Bobylev, Bronnikova \&
Shakht, 2004; I/285 CDS) for the faint stars programme (KSZ), the
NPM1 (Klemola et al., 1987; III/199 CDS) for the Lick Northern
Proper Motion, the SPM2 (Platais et al., 1998; III/277 CDS), for
the Yale Southern Proper Motion. The maximal number of stars 287
thousand is contained in the SPM2 catalogue, while the maximal
number of galaxies, approximately 70 thousand, is in the NPM1
catalogue. The GPM, PUL2 and NPM1 catalogues cover the northern
sky and partially the southern one, and the SPM2 catalogue covers
about one third of the southern sky. The random error of proper
motions in these catalogues depends on stellar magnitude and
varies from 3 to 10 mas/yr, while the accuracy of the absolute
calibration is 2--5 mas/yr.

The above-mentioned catalogues of absolute proper motions are very
important for astrometry, since they allow the local coordinate
system to be implemented, which does not rotate with respect to
galaxies. The global quasi-inertial coordinate system can be
established through the catalogue of absolute proper motions of
stars covering the whole sky. The data of these catalogues play
the principal role in determining kinematic parameters of the
Galaxy, for example, in the framework of the model by
Ogorodnikov-Milne. It is worth noting that this model provides the
most adequate parameters, on conditions that the proper motions
representing the whole celestial sphere are used.

As was mentioned in Paper I, the XPM catalogue contains
approximately 280 million absolute proper motions of stars and
covers the whole celestial sphere, excluding a narrow zone near
the galactic equator  within the stellar magnitude range from
$11^{m}<B<20^{m}$. The random error of its proper motions depends
on stellar magnitude and lies within 3--10 mas/yr, the error of
absolute calibration in the northern hemisphere is approximately
0.3 mas/yr, and of the order of 1 mas/yr in the southern one.
Creation of this catalogue is based mainly on the three most
important procedures:

\begin{enumerate}
\item cross-identification, which allows to identify and compare
objects in the USNO-A2.0 and 2MASS catalogues;

\item elimination of systematic errors in positions of the
USNO-A2.0 objects with the use of the median filter;

\item derivation of the absolute proper motions of stars.
\end{enumerate}

Evidently, the cross-identification procedure is crucial in the
set listed above, since it determines all other procedures and the
resulting accuracy of the absolute proper motions. It has been
noted in Paper I that the cross-identification procedure mentioned
above is not, strictly speaking, an actual cross-identification,
but it is rather an association that can result in false
identifications. This leads in turn to forming false position
differences for stars and galaxies. Thus, the values of function
$F(\alpha,\delta)$ obtained with the median filter (see Paper I)
will be burdened with the errors, which will inevitably result in
erroneous proper motions. Therefore, most of attention must be
given to the cross-identification procedure.

In the XPM-1.0 version we used a somewhat improved version of the
cross-identification procedure as compared to the previous version
of XPM described in Paper I. It was only for this procedure that
proper motions from the USNO-B1 catalogue (Monet et al., 2003)
were involved. This has made it possible to carry out intersection
of three catalogues --- USNO-1, USNO-2.0 (Monet et al., 1998) and
2MASS (Skrutskie et al., 2006) using a circular search window of
1.5 arcsec in dimension. Moreover, the high-precision photometric
data of 2MASS were used to calculate the USNO-2.0 magnitudes,
which were compared to their original values in selection the
objects within the circular 1.5-arcsec search window. This is
described in details in Section 2. There is no simple test at this
stage, which would allow to quantitatively estimate the
improvement of the catalogue properties. It is caused first of all
by the absence of the accuracy estimates for individual positions
of stars in the initial catalogues. Nevertheless, we believe that
using of the improved version of the cross-identification
procedure results in a decrease of random errors in position
differences, in some broadening of the stellar magnitude range, as
well as in improvement of linking to extragalactic objects.

According to the idea of creating the most comprehensive
catalogue, we derive the proper motion of stars in the fields,
which are not supplied by the number of galaxies sufficient for
absolute calibration. If the number of galaxies in a particular
field is not sufficient for absolute calibration, we do not
exclude this field from consideration. Unlike previous version of
the XPM catalogue, we use a special absolute calibration procedure
in these fields. To do this, the parameters of reduction model of
absolute calibration inside every field with an insufficient
number of galaxies were calculated by a two-dimensional
interpolation between the corresponding values from the
neighboring fields. We use the term quasi-absolute calibration for
the procedure of estimating proper motions in such fields, and
describe it qualitatively in Section 3. Thus, after application of
the procedures described above, each field of the total 1431 will
eventually contain the absolute or quasi-absolute proper motions
of stars.

Although using of the median filter noticeably decreases the
geometrical distortions in positions of the USNO-A2.0 objects, the
photometric (magnitude-dependent) distortions in their positions
remain unchanged after the median filter is applied. Therefore, we
undertake efforts to eliminate the magnitude equation in the
XPM-1.0 catalogue mainly in the faint end of the range of stellar
magnitudes. Section 4 is dedicated to the search and analysis of
the magnitude equation in the catalogue.

Section 5 is dedicated to comparison of the XPM-1.0 catalogue with
UCAC-2.0 (Zacharias et al., 2004) and UCAC-3.0. The UCAC-3.0
catalogue ({\tt http://www.usno.navy.mil/usno/astrometry}) is the
only one, which can be used  to compare proper motions over the
whole celestial sphere. Though such a comparison is not correct
enough because of the fact that the UCAC-3.0 proper motions are in
the International Celestial Reference System (ICRS) (Arias et al.,
1995), the qualities of both catalogues can be estimated.

This version of the XPM catalogue contain approximately 280
million objects covering the whole sky in the magnitude range
$10^m<B<22^m$. Their positions and absolute proper motions are
presented, as well as the standard $J,H,K,B$ and $R$ magnitudes
taken from 2MASS and USNO-2.0. For those stars from the XPM-1.0
catalogue which resulted from intersection of the USNO-B1, 2MASS
and USNO-A2.0 catalogues, the magnitudes of USNO-B1 are also
included. It should be emphasized that the XPM-1.0 catalogue is
obtained using the data of two ground-based catalogues, --- 2MASS
and USNO-A2.0, --- and contains absolute proper motions. Positions
in XPM-1.0 are given on the ICRS, since the stars from the 2MASS
catalogue are given in this system.

\section{On the cross-identification}
A preliminary investigation has shown that the XPM catalogue
contains relatively many misidentified stars, especially, at the
faint end of the stellar magnitude range. It is small wonder,
since in the fields with a high star density in the circular
window with the radius  of 3.5 arcsec may fall onto several
objects.  These false identifications have led to the smearing of
systematic coordinate differences on which the construction of the
median filter was based to eliminate the systematic errors in the
USNO-A2.0 catalogue, and ultimately to errors in the absolute
proper motions. In this article we describe a slightly different
approach, which has provided a more reliable cross-identification
of stars and galaxies contained in the USNO-A2.0 and the 2MASS
catalogues. The essence of this approach consists in diminishing
the window radius to 1.5 arcsec and in comparing  the calculated
and original catalogue magnitudes in this window. Thus, this
approach greatly increases the probability of the correct
identification of objects in catalogues.

\subsection{Coordinate identification}
To implementation this approach first of all we have found evident
systematic offsets between the positions of objects in USNO-A2.0
and 2MASS for the southern and northern hemispheres, separately.
The systematic difference between the positions of galaxies in the
USNO-A2.0 and 2MASS can reach up to 2--3 arcsec, which is
consistent with research USNO-A2.0 by (Assafin et al., 2001).
After the exclusion of systematic coordinate offsets, we attract
the proper motions of stars from purified USNO-B1 catalogue
(Barron et al., 2008).

The procedure for the identification of stars in the circular
search window with the 1.5 arcsec radius consists of two steps.
First, we match the objects of the USNO-A2.0 and USNO-B1
catalogues, using the encoding of  surveys and fields as given in
the description of the USNO-B1 format. Thus a subset of objects
are selected from the USNO-B1 catalogue that were used to compile
the USNO-A2.0 one.

Then we reduce the positions of stars with proper motions from the
USNO-B1 catalogue to the epoch of a particular field of the
USNO-A2.0. For stars with no proper motions we use the positions
from USNO-B1, which are formally given as referred to the epoch
J2000, but actually they are referred to the epoch equal to the
average of epochs, the used surveys are referred to.
Unfortunately, only about 285 million out of one billion USNO-B1
objects, have the proper motions, and of these, only about 4
million stars have the proper motions, exceeding 30 milliarcsec
per year. For other objects in the USNO-B1 catalogue the zero
proper motions are given. The differences between the positions of
these objects in both catalogues due to their proper motions do
rarely exceed 1 or 1.5 arcsec, since for 20 through 25 years, i.
e. for the difference between  the mean epoch and the first one,
the stars are displaced no more than by 1 through 1.5 arcsec even
when their  proper motions are about 60 through 75 milliarcsec per
year. Thus, we use for the identification of objects in the search
window with the radius of 1 through 1.5 arcsec not only the stars
with proper motions taken from the USNO-B1 catalogue, but also
those with the ``zero proper motions'' taken from the same
catalogue. Since by deriving the positions of the USNO-B1 objects
the same surveys, as for those of the USNO-A2.0 catalogue among
others were used,  it is obvious that the systematic differences
between the USNO-B1 and USNO-A2.0 star positions are strongly
correlated, so that their values seldom exceed 0.75 arcsec.
Therefore, the uncertainties of positions of stars in the USNO-B1
catalogue due to the random and systematic errors of the positions
and proper  motions, are equal to 0.75 through 1.00 arcsec even
for the  epochs falling into the 1950s.

For the final cross-identification of objects USNO-A2.0 and
USNO-B1 we have used the search window with the 1.5 arcsec radius.
In addition, we have compared the stellar magnitudes of the
USNO-B1 stars and those of the  USNO-A2.0 ones on the entire range
of stellar magnitudes, besides making use of the coordinate search
window. Thus, we have got the intersection of two sets in the form
of a list of the USNO-A2.0 and USNO-B1 objects identified in the
search window with the 1.5 arcsec radius. As the next step, we
identify the USNO-B1 objects from the resulting list and the 2MASS
objects. As already mentioned, the positions in the USNO-B1
catalogue are formally given as referred to the epoch J2000, with
the exception of stars with the ``zero proper motion''. The epochs
of the positions of these stars are the average epochs of the ones
of the surveys used. As shown above, for these stars the
displacement for the 25 years does not exceed 1.5 arcsec. The
differences between the coordinates of stars and galaxies in the
2MASS and USNO-B1 catalogues are basically originated by the
systematic and random errors of these catalogues and do not exceed
0.75 arcsec. Therefore, for the cross-identification of objects,
and in the present case, too, we have used the search window with
the 1.5 arcsec radius.

\subsection{Photometric identification}
As mentioned in the Paper1, we were not able to perform the
full-fledged cross-identification, so that we restrict ourselves
to the positional association only. But it is clear that the
coordinate criterion taken alone is not sufficient for identifying
the stars, and, particularly, those having been observed in the
optical and the near infrared range.  Therefore it is necessary to
apply an additional criterion to identifying the USNO-A2.0 and the
2MASS objects. The photometric criterion is commonly being used as
such a criterion, but it is impossible to directly compare  the
USNO-A2.0 stellar magnitudes and the 2MASS ones. However, when
analyzing the previous version of the XPM catalogue we have found
out that the photometry of the USNO-A2.0 catalogue for the
northern hemisphere is different from that for the southern one.
For example, the average magnitude $B$ end $R$ of galaxies in the
northern and southern hemispheres differs systematically by about
2 magnitudes, and for stars this difference is about 0.5--2
magnitudes. It is  difficult to  use the  unified photometric
criterion for identifying the USNO-A2.0 and 2MASS objects because
of these facts.

The solution of two tasks appeared to be necessary to  resolve
this problem. First, the magnitudes of all objects should be given
in a common system, even if not in the entirely accurate
photometric one. As such the  system  given by the magnitudes of
objects of the northern hemisphere of the USNO-A2.0 catalogue was
chosen. After that, a method for determination of the $B$ and $R$
stellar magnitudes of these objects should be found, which  is
based on their $J,H$ and $K$ magnitudes from 2MASS catalogue.

To solve the first task we have constructed the relationships
between the $B$ and $R$ stellar magnitudes of the previous version
of the XPM catalogue and the $J$ magnitude of the 2MASS catalogue
separately for the northern hemisphere, the photometry of which
being taken as the  basic one. By using similar relationships
obtained in each particular USNO-A2.0 field  the $B$ and $R$
magnitudes of all objects in this field were reduced to the basic
photometric system.

To solve the second task, we applied the method for calculating
the stellar magnitudes of USNO-A2.0 using a more accurate
photometry described by Sesar et al. (2006). In our case the
reference stellar magnitudes were those of the 2MASS catalogue. By
use of the data for the entire celestial sphere as given by the
previous version of the XPM catalogue, the functions $f_{1}$  and
$f_{2}$ have been determined  separately for stars and galaxies
from the following equations:
\[
{B_{XPM}=J_{2MASS} +f_{1}(J_{2MASS}-K_{2MASS})}
 \]
 \[
{R_{XPM}=J_{2MASS} +f_{2}(J_{2MASS}-K_{2MASS})}
\]
To obtain a sufficiently detailed behavior of $(B-J)$ against
$(J-K)$ from the data of the first version of the XPM catalogue,
the full range $(J-K)$ was divided into sub-ranges of 0.25 mag in
width. The average value of $(B-J)$ of each sub-range was
calculated. This dependence was approximated by a 9-th power
polynomial (Fig. 1). The behavior of the polynomial at the edges
was fixed by cutting the marginal points of $(J-K)$ range.

In the new procedure for identifying the objects the obtained
functions  $f_{1}(J_{2MASS}-K_{2MASS})$ and
$f_{2}(J_{2MASS}-K_{2MASS})$ were used to calculate the stellar
magnitudes ${B_{2MASS}}$ and ${R_{2MASS}}$ of the 2MASS catalogue
objects, falling into the circular coordinate window. To choose
between the candidates caught in the circular window the following
conditions were used
\[
{B_{USNO-A2.0} - B_{2MASS} < 1.00^m}
\]
\[
 {R_{USNO-A2.0} - R_{2MASS} < 0.75^m}
\]
In addition, for the analysis of the signs of the CI (color
indices) $(B-R)$ and $(J-H)$ we applied the procedure which allows
a more reliable selection of stars. The basis for such a procedure
is  constituted by a simplifying assumption that in most cases the
intensity distribution in the star's spectrum is the unimodal one.
The correct identification of the stars caught into the search
window is performed in accordance with this assumption only in 3
cases:

\begin{enumerate}
 \item In the first case the CI $(B-R)$ and $(R-J)$ $>$ 0 which
corresponds to the monotonic increase of the intensity in the
range from the blue to the infrared part of the spectrum.
 \item In the second case the CI $(B-R)$ and $(R-J)$ $<$ 0 which corresponds
to the monotonic decrease of the intensity in the same range of
the spectrum.
 \item In the third case CI $(B-R)$ $>$ 0, but CI
$(J-H)$ $<$ 0 which corresponds to the intensity maximum situated
between the B and H magnitudes.
\end{enumerate}

In accordance with the sign of the color index, we place either
the USNO-A2.0 star or the 2MASS one in the center of the search
window. This allows not to consider those objects which may be
contained in one catalogue only due to their intensity
distribution in the spectrum, i. e. either in the optical
catalogue or in the infrared one. It should be noted that we are
not aiming at improvement of the photometry of the USNO-A2.0
catalogues. Our goal is to be able to compare the original
USNO-A2.0 magnitudes  with the magnitude values calculated using
the photometry of the 2MASS catalogue, in addition to identifying
objects in the coordinate window. And finally, one more remark. In
the highly-dense fields containing more than 500 thousand objects,
the cross-identification between the USNO-A.2.0 and the 2MASS
objects was carried out without using the proper motion of
USNO-B1, but with using the photometric cross-identification. This
is due to the fact that when performing the identification of
objects in the field with the object number not exceeding 500
thousand, the rate of the identified USNO-B1 and 2MASS objects is
more than 90\%, whereas in denser fields this rate dropped down to
45--50\%.

After the cross-identification the approximating function
$F(\alpha,\delta)$ (see Paper I) inside each field were derived by
using the coordinate differences of all the star pairs. In
addition, the coordinate differences inside each field were
approximated by rough linear relationships which we have used for
the cross-identification of galaxies.

The procedure of the cross-identification of galaxies is crucial
for the absolute calibration. The reliable cross-identification of
galaxies ensures a valid reduction of the observed proper motions
of stars to a coordinate system that does not rotate in the space.
On the other hand, among the extended sources from the XSC
catalogue there are not only extragalactic objects but also the
objects in the Milky Way which have the proper motions. Therefore,
the procedures of the cross-identification for galaxies and for
stars were performed separately. Obviously, after  subtraction of
the approximating functions $F(\alpha,\delta)$ from the initial
function $\Delta P(\alpha,\delta)$, i.e. after reducing the
coordinates of all the USNO-A2.0 objects to the 2MASS coordinate
system, the revised coordinate differences of all stars on average
are equal to zero, whereas the revised coordinate differences of
the galaxies on the average have a value that approximately equal
to the average proper motion in this field, but with opposite
sign. To perform the correct identification of galaxies in the
search window with a radius of about 0.8 arcsec, their revised
coordinate differences were corrected by using the linear
relationship mentioned in the preceding paragraph. Next, we
identify the XSC and USNO-A2.0 objects in the circular window of
the 0.8 arcsec radius only, since at this step the position
differences between the USNO-A2.0 and the XSC galaxies caused by
only their position random errors. Theoretically, this will lead
to eliminating the extended objects with non-zero proper motions
from consideration.

Thus, the applied approach allows to improve the
cross-identification between the USNO-A.2.0 and the 2MASS objects.
Owing to these procedures used for the cross-identifications, the
contamination rate of the spurious entries in the XPM-1.0
catalogue was decreased visibly, and the quality of linking to
extragalactic objects was improved. Operations for linking to the
extragalactic objects and deriving the absolute proper motions do
not differ fundamentally from those described in the preceding
article.

\begin{figure}
\begin{center}
\includegraphics[width = 78mm,]{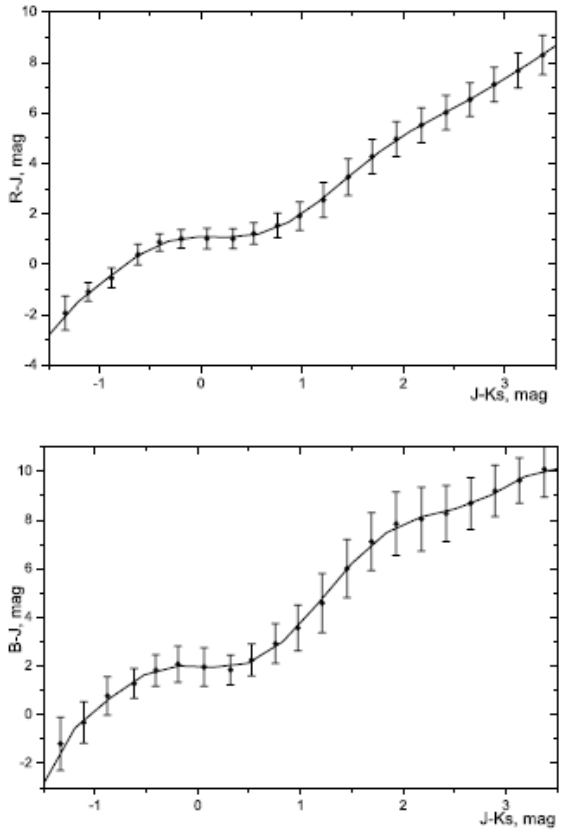}
\caption{The fitting  curves for $f_{1}$ and  $f_{2}$ functions
for calculation of $R$ and $B$ magnitudes and distribution of the
means and the standard deviations.}
\end{center}
\end{figure}

\section{Quasi-absolute calibration}

The procedure of the  absolute calibration was described  in
detail in the preceding article. Here we specify only the fields
for which this procedure is not entirely correct. Because the
galaxies are practically invisible in the zone of avoidance,
particularly, in the direction to the galactic center, the
absolute calibration has not been implemented in the fields which
cover this zone or which contain less than of 25 galaxies.
However, it is well known that this particular zone is of a great
interest for astrophysics and stellar astronomy. Moreover, XPM-1.0
contains the fields in which the distribution of galaxies does not
appreciably symmetrically about the center. If the number of
galaxies in these fields was less than 100, the absolute
calibration also was not performed. Therefore, we applied a
procedure called by us the quasi-absolute calibration to these
fields. The essence of this procedure is as follows. First, to
fulfill the absolute calibration in every field with a sufficient
number of galaxies, we determined the parameters of reduction
model
 $\phi(\alpha, \delta)=\Delta P_{gal} (\alpha, \delta)-F(\alpha, \delta)$
 (see Paper I) from the coordinate differences of
the galaxies. The function $\phi(\alpha,\delta)$ represents
evidently a distribution of the mean proper motion of stars in the
field in question taken with an opposite sign. To derive
quasi-absolute proper motions in the field where the absolute
calibration is impossible, we obtained the parameters of the
reduction model for this field by interpolation of the
$\phi(\alpha,\delta)$ values from a surrounding area of the 2 by 2
field having applied several iterations.

\begin{figure}
\begin{center}
\includegraphics[width = 78mm,]{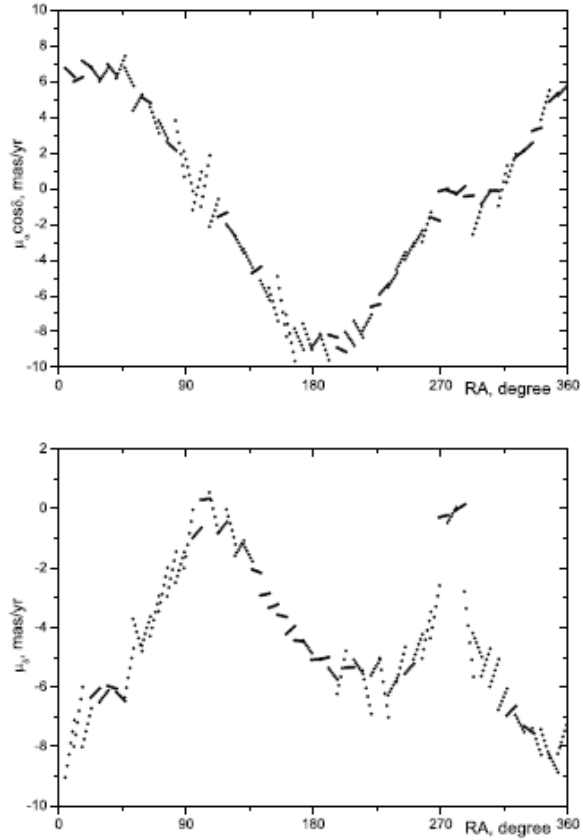}
\caption{The mean proper motions as functions of the coordinates
in fields having  declinations approximately from $-7.5^\circ$ to
$-2.5^\circ$ and located in the band of right ascensions from
$0^\circ$ to $360^\circ$. Each field in Fig. 2 is represented by
six points of averaged proper motion.}
\end{center}
\end{figure}

In this case, we assume that the motion of stars in the sky can be
described by a continuously differentiable function. For example,
in one-dimensional case, the fields that contain no galaxies are
seen in Fig. 2 near RA$=270^\circ$. The mean proper motion in
these fields is significantly different from that of the
neighboring fields. Therefore we obtained the mean proper motion
for the fields that contain no galaxies by interpolation of the
corresponding values from the neighboring fields. The 67 fields
(45 in the southern hemisphere and 22 in the northern hemisphere)
in which the quasi-absolute calibration procedure had been carried
out were marked by a special flag in the catalogue. This approach
also allows (see Fig. 2) to inspect visually the absolute
calibration validity. The rest of procedures for these fields in
principle do not differ from the described previously.
Unfortunately, there is no possibility to test the method at this
stage, so we are planning to do this in our future investigations.
To approximately estimate the uncertainty of the quasi-absolute
calibration, we used the value that does not exceed a
half-difference of the mean proper motions from the neighboring
fields.

\section{The magnitude equation}
Under the term ``the magnitude equation'' the unwanted correlation
between the measured position of the star image and its magnitude
is commonly understood. The main causes of this phenomenon are
assumed to be the optical misalignment, optical aberrations and
the inevitable errors of the telescope guiding. They lead to the
asymmetry of stellar profile and dissimilar to a point spread
function and combined with the nonlinear response of the emulsion
they lead to the differing profiles of images of stars with
different magnitude. As a result there is a systematic bias of the
measured centers of stellar images depending on the apparent
brightness. The magnitude equation in the proper motions of the
XPM catalogue is a result of the difference of the magnitude
equations present in the positions of USNO-A2.0 and 2MASS
catalogues. As to the magnitude equation in the 2MASS catalogue,
there is no information but we hope that if it would be available,
the magnitude equation would be not very large, because the
observations were made with the CCD detectors. Concerning the
magnitude equation of 2MASS catalogue it is reasonable to assume
that it caused by Charge Transfer Efficiency (CTE) effects and can
induce a systematic errors of the position centroids CCD but we
hope that they is not very significant.

The USNO-A2.0 catalogue had been compiled on the basis of three
photographic surveys, i. e. POSS-I, ESO/SERC $J$ and ESO/SERC $R$.
As it is well known, the POSS-I survey covers the whole northern
sky and the part of the southern sky from 0$\degr$ to $-30\degr$
in declination. Our experience based on the work  with scanned
images of the photographic plates POSS-I survey indicates that the
magnitude equation present in the O and E plates in the range of
Tycho-2 stellar magnitudes is negligible (Fedorov \& Myznikov,
2006).

In the southern hemisphere the surveys were made with two Schmidt
telescopes. One of them was located in Australia  $(\varphi =
-31\degr 27', \lambda = 149\degr 07')$. With this telescope 606
blue plates in the declination range from $-20\degr$ to $-90\degr$
were taken in 1975--1987 with the blue filter GG 395 (3950--5400
Angstroms). The corresponding plates with the filter RG 630
(6300--6900 Angstroms) were taken in 1978--1990 with the Schmidt
telescope of the La Silla Observatory in Chile $(\varphi =
-29\degr 15', \lambda = 70\degr 44')$.

Thus, it is clear that the magnitude equation present in each of
these surveys is originated by the causes which are intrinsic to a
specific survey only, and ideally it should be studied separately.
However, there is no such a possibility, because the USNO-A2.0
catalogue contains the averaged coordinate values assigned to the
mean epoch of the blue and red plates. For the northern hemisphere
and for the part of the southern one (up to $-17.5\degr$ in
declination), the observations were made with the red and blue
filters during one night with the same telescope, and the mean
epochs of the red and blue plates are essentially identical. For
the southern hemisphere the observations were made under different
conditions, with different telescopes and with different filters.
Obviously, the magnitude equations present in these two parts of
the catalogue should be different. Therefore, the magnitude
equation should be examined in each specific field in order to
most reliably eliminate it.

\subsection{Influence of the magnitude equation on the  absolute calibration}
For an arbitrary field of the XPM-1.0 catalogue the proper motion
of any star, depending on the coordinates may be represented by
the expression:
\[
\mu(\alpha, \delta)_{i}=\mu_{true}(\alpha, \delta)_{i} +
\varphi(\alpha, \delta)_{i} +f[m_{i}(\alpha, \delta )],
\]
where $\mu_{true}(\alpha, \delta)_{i}$  is the true proper motion
of any arbitrary star, $\varphi(\alpha, \delta)_{i}$  is the
coordinate systematic error caused by systematic coordinate errors
in both catalogues, being inherent to all objects in the field
given, and $ f[m_{i}(\alpha, \delta )]$ --- is the systematic
photometric error caused due to different displacements of the
photometric centers of stars with various stellar magnitudes, i.e.
the magnitude equation. The bright stars are shifted from the true
center stronger than the faint ones. As a result, a fictitious
proper motion is arisen with a greater value for the bright stars
than for the faint ones. When the coordinate dependence of the
proper motions of the field stars is approximated by a linear
relationship, we obtain the coordinate dependence of the mean true
proper motion of stars distorted by the mean coordinate error and
the mean photometric one:
\[
\langle\mu^{S}(\alpha,
\delta)\rangle=\langle\mu_{true}^{S}(\alpha,
\delta)\rangle+\langle\varphi^{S}(\alpha, \delta)\rangle+\langle\
f[m^{S}(\alpha, \delta )]\rangle.
\]
The absolute calibration of the proper motions of stars involves
the use of formal mean proper motions of galaxies:
\[
\langle\mu^{G}(\alpha, \delta)\rangle=\langle\varphi^{G}(\alpha,
\delta)\rangle+\langle\ f[m^{G}(\alpha, \delta )]\rangle.
\]
Because the true proper motions of galaxies are equal to zero and
the coordinate mean errors $\langle\varphi^{S}(\alpha,
\delta)\rangle$ end $\langle\varphi^{G}(\alpha, \delta)\rangle$
are differing only randomly as a result of a random sampling, the
procedure of the  absolute calibration is the following:
\[
\langle\mu^{ABS}(\alpha, \delta)\rangle = \langle\mu^{S}(\alpha,
\delta)\rangle-\langle\mu^{G}(\alpha, \delta)\rangle;
\]
\[
\langle\mu^{ABS}(\alpha,
\delta)\rangle=\langle\mu_{true}^{S}(\alpha,
\delta)\rangle+\langle\ f[m^{S}(\alpha, \delta \rangle-\langle\
f[m^{G}(\alpha, \delta )]\rangle.
\]
Many stars with different proper motions and different magnitudes
are contained in each range of coordinates (right ascensions and
declinations). But the faint stars make the most large
contribution to the value of
\[
\langle\ f[m^{S}(\alpha, \delta)]\rangle = \frac{1}{N}\sum
f[m^{S}_{i}(\alpha, \delta)],
\]
since they are the most numerous in each sub-range. In other
words, we may say that the average value of the magnitude equation
in the field will be approximately equal to the  magnitude
equation value for the mean stellar magnitude of this field. This
means that the contribution of the average magnitude equation to
the coordinate dependence of the average proper motion is
practically zero. Similarly, the faint galaxies the magnitude
equation  of which is practically also equal to zero, make the
main contribution to the value of
\[
\langle\ f[m^{G}(\alpha, \delta)]\rangle = \frac{1}{N}\sum
f[m^{G}_{i}(\alpha, \delta )].
\]
Thus, we can conclude that the magnitude equation almost does not
influence the process of the absolute calibration and remains
unchanged in the absolute proper motions of the XPM-1.0 catalogue.

\begin{figure*}
 \includegraphics[width = 155mm]{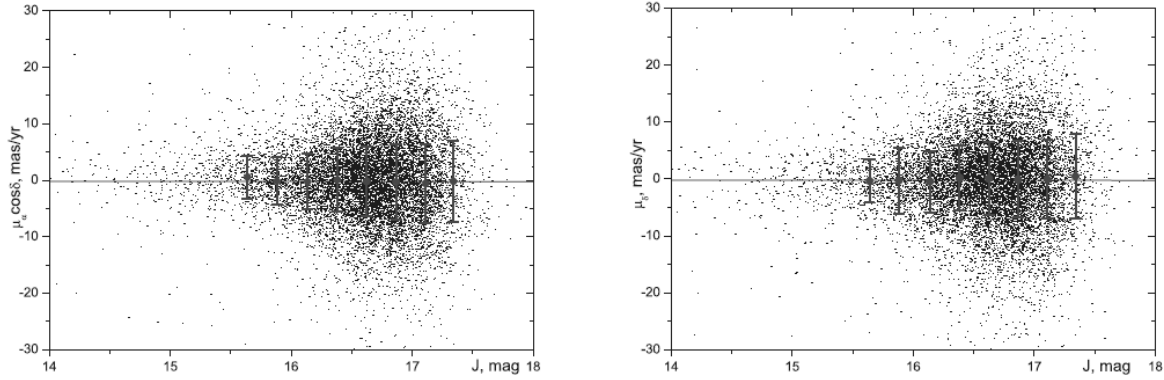}
\caption{Scatter of individual formal proper motions
$\mu_{\alpha}\cos\delta$ (left) and $\mu_{\delta}$ (right) DR5
quasars as a function of magnitude $J$. The red solid circles and
lines show the mean values and standard deviations.}
\end{figure*}
\begin{figure*}
 \includegraphics[width = 155mm]{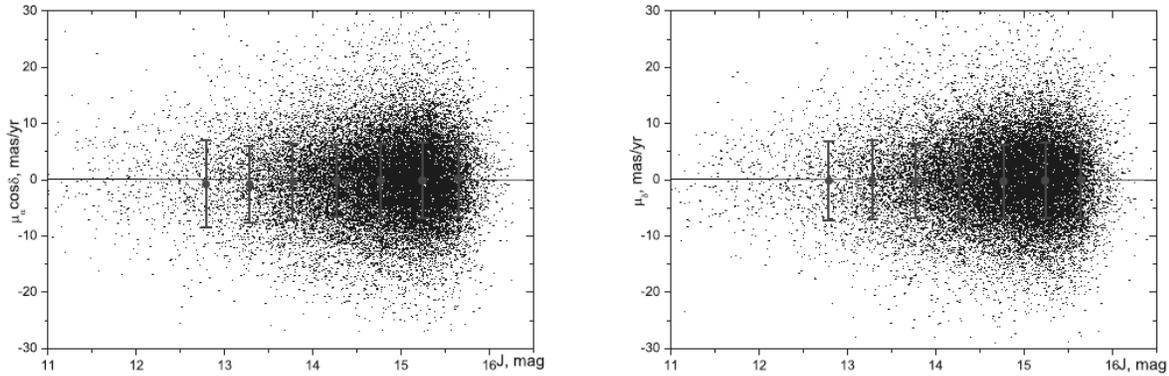}
\caption{Scatter of individual formal proper motions
$\mu_{\alpha}\cos\delta$ (left) and $\mu_{\delta}$ (right)
galaxies as a function of magnitude $J$. The red solid circles and
lines show the mean values and standard deviations.}
\end{figure*}

\subsection{The magnitude equation in the faint part}
To study the magnitude equation in the faint range of stellar
magnitudes we have used the quasars. The profiles of their images
are very close to the stellar ones, which are usually constituting
the basis for correction of the magnitude equation. Since the
proper motions of quasars are equal to zero, it should be
reasonable to  interpret any magnitude dependence of their formal
proper motions as the magnitude equation. Since the quasars were
not used for  the absolute calibration by the derivation of proper
motions of the XPM-1.0 catalogue, their absolute proper motions
were derived exactly in the same way as for stars. Therefore,
their own formal proper motion may well be used to verify the
existence of the magnitude equation in the faint end of the range
of stellar magnitudes. Unfortunately, at present the most complete
catalogue of quasar positions, i.e. the catalogue SDSS DR5
(Schneider et al., 2007) covers only a part of the celestial
sphere and, therefore, it is not possible to investigate  the
magnitude equation throughout the XPM-1.0 catalogue.
Approximately, 12 thousand quasars from the DR5 were found in the
XPM-1.0 catalogue.

The formal proper motions of quasars as  functions of  the stellar
magnitude are shown on Fig. 3. It is obvious that there is no
dependence, and the mean value of formal proper motions are 0.12
and $-0.24$ mas/yr of $\mu_{\alpha}\cos\delta$  and
$\mu_{\delta}$, respectively. The standard deviations of
$\mu_{\alpha}\cos\delta$ and $\mu_{\delta}$ are estimated to be
approximately 3.8 through 7.4 mas/yr. Thus we may conclude that in
the right ascension and declination areas  of the XPM-1.0
catalogue, intersecting with the DR5, the magnitude equation is
absent in the ranges from about 15 to 20 stellar magnitude. The
formal proper motions of galaxies (taken as the residual
discrepancies in the coordinates of galaxies divided through the
epoch differences) versus the stellar magnitude considered in the
same overlapping zones are shown on Fig. 4. As it is seen from the
figures, there is no distinction between these relationships, so
that we can use the galaxies in each USNO-A2.0 field for
elimination of the magnitude equation in the faint end of the
range of stellar magnitudes.

\subsection{Analysis of the magnitude equation in the bright star range of the XPM-1.0 catalogue}
To study the magnitude equation in the bright end of the range of
stellar magnitudes we used the TYCHO-2 catalogue and the UCAC-2.0
one (H\o{}g et al., 2000; Zacharias et al., 2004). We assume that
there are no magnitude equations in the TYCHO-2 and the UCAC-2.0
catalogues. In theory the difference between the proper motions of
stars from these catalogues and of those from the XPM-1.0
catalogue can be represented as:
\[
\mu^{ABS}(\alpha, \delta,m) - \mu^{kat}(\alpha,
\delta)=\mu^{ABS}_{true}(\alpha, \delta)-\mu^{kat}_{true}(\alpha,
\delta)
\]
\[
+ \Delta \mu(m) + \Delta \mu_{0}(\alpha_{field}, \delta_{field})
\]
where $\Delta \mu(m)$ --- depends on the magnitude, but does not
depend on the coordinates, and $\Delta \mu_{0}(\alpha_{field},
\delta_{field})$ --- does not depend on the magnitude but depends
only on the coordinates of a particular field and presumably is
caused by the  differences of proper motion systems of both XPM
and TYCHO-2 catalogues. If we construct the dependence of the
proper motion differences versus the magnitude  in every field
\[
\mu^{ABS}(\alpha, \delta,m) - \mu^{kat}(\alpha, \delta)=\Delta
\mu(m)+\Delta \mu_{0}(\alpha_{field}, \delta_{field}),
\]
we can determine the form of the dependence in the range of the
TYCHO-2 and UCAC-2.0 stellar magnitudes only. As can be seen, by
the use of the proper motions of the TYCHO-2 end UCAC-2.0 stars
the magnitude equation in the XPM-1.0 catalogue may be determined
up to a constant only. Thus the elimination of the magnitude
equation by using the  TYCHO-2 proper motions means that the
system of the proper motions of the XPM-1.0 catalogue ceases to be
an independent realization in the bright part, being linked to the
system of proper motions of the HIPPARCOS (Perryman et al., 1997;
Kovalevsky et al., 1997) via of TYCHO-2 catalogue stars.
Therefore, we have left the magnitude equation in the bright part
of the XPM-1.0 catalogue for a while unchanged.

\begin{figure*}
 \includegraphics[width = 155mm]{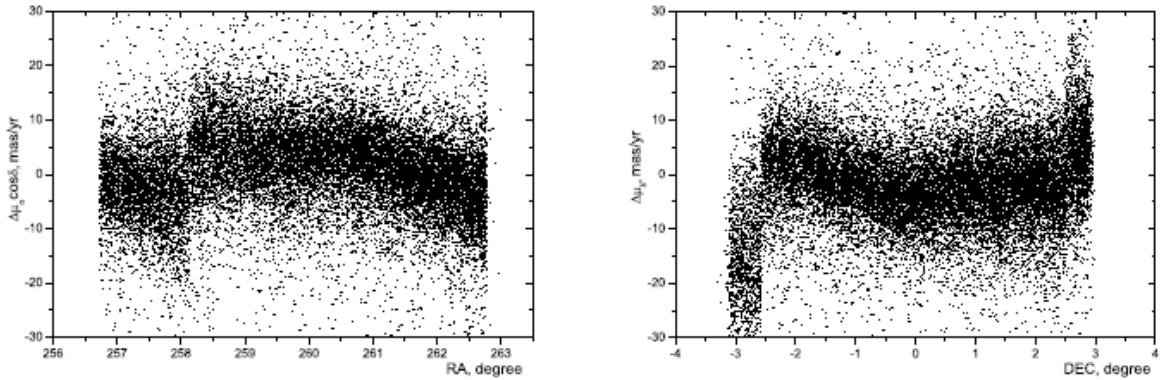}
\caption{The individual differences of proper motions of stars
(XPM1.0--UCAC3.0) in selected field as a function of RA and Dec.}
\end{figure*}
\begin{figure*}
 \includegraphics[width = 155mm]{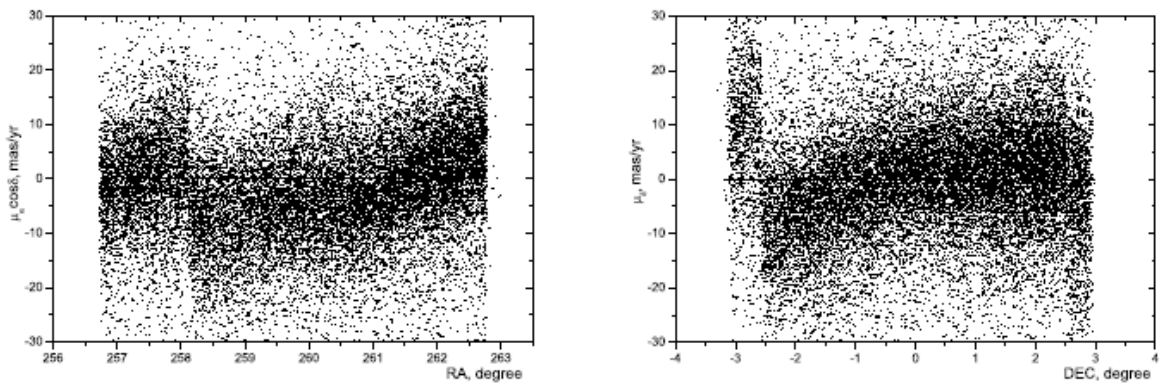}
\caption{The proper motions of UCAC3.0 stars in selected field as
a function of RA and Dec.}
\end{figure*}

\section{Comparison of XPM-1.0 with other catalogues of proper motions}
After consideration of the magnitude equation of the XPM-1.0
catalogue we have compared it with other catalogues with the  aim
to get an idea about the consistency of the absolute proper
motions of stars with the relative ones obtained in the
HIPPARCOS/TYCHO-2  system. Today there are several catalogues of
proper motions of stars, but by no means all of them could be used
for this comparison. Some of these catalogues contain the absolute
proper motions and cover the northern or southern hemisphere only,
such as NPM1 (Klemola et al., 1987), and SPM2 (Platais et al.,
1998). Though other catalogues cover  partially or almost the
entire celestial sphere they contain, however, the relative proper
motions of stars only (Girard et al., 2004; Hanson et al., 2004;
Zacharias et al., 2004; Monet et al., 2003). Prima facie the
USNO-B1 and the UCAC-2.0, 3.0 catalogues are the most suitable
ones for this purpose.

The USNO-B1.0  catalogue, covering the entire sky up to 21
magnitude, and containing positions, proper motions, and other
data, provides  the astrometric accuracy of 0.2 arcsec at the
epoch J2000. The proper motions given in the catalogue are
relative. As noted earlier, despite the fact that in the catalogue
the positions of about one billion stars are given, the proper
motions are given for 285 million objects only. The proper motions
for the remaining approximately 760 million stars in the catalogue
are equal to zero. This fact greatly complicates the
identification of stars in the catalogues and the direct
comparison of their proper motions. In addition, the catalogue
contains a great many (tens of millions) of artifacts (Barron et
al., 2008). These facts compelled us to abandon the use of the
USNO-B1.0 catalogue for comparison with the XPM-1.0.

\begin{figure*}
 \includegraphics[width = 155mm]{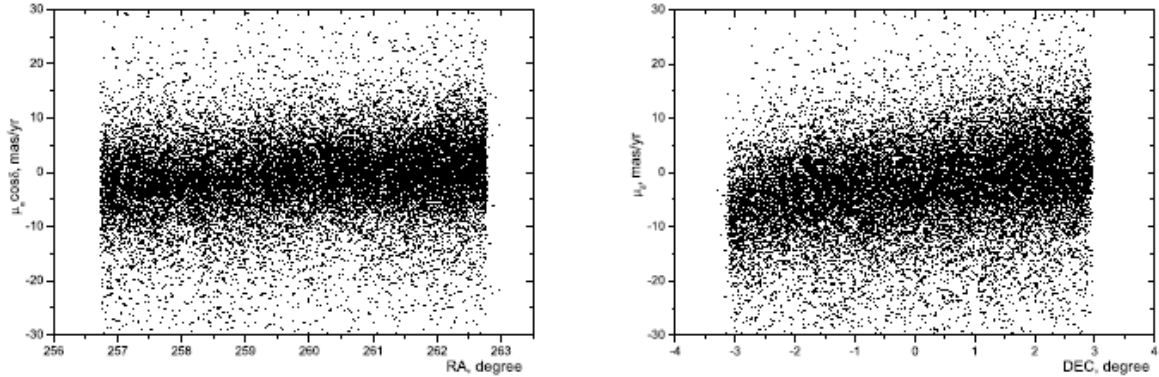}
\caption{The proper motions of XPM1.0 stars in selected field as a
function of RA and Dec.}
\end{figure*}

UCAC-2.0 is previous version of catalogue UCAC-3.0. The UCAC-3.0
is the dense astrometric catalogue  of the high precision,
containing 100,766,420 stars, covering the entire sky. The errors
of its positions are from 15 to 20 milliarcsec for the stars in
the range from 10 to 14 R magnitude and about 70 milliarcsec for
other stars up to 16 magnitude. The errors of proper motions of
bright stars (up to 12 magnitude) are in the range of 1 through 3
milliarcsec per year. For the fainter stars, the positions of
which were taken from the SPM the typical errors are estimated to
be approximately 2 through 3 milliarcsec per year, and for the
data taken from the early epoch of SuperCOSMOS, the typical error
is 6 through 8 milliarcsec per year. The positions and proper
motions of stars are given in the ICRS for the epoch J2000.0. The
comparison of the proper motions in star catalogues was carried
out by following two simple ways, namely:
\begin{enumerate}
 \item The individual differences of proper motions of stars in the
selected fields were calculated.

 \item The systematic differences of proper motions as well as
their dispersions, depending on the magnitude were computed.
\end{enumerate}
To compare the proper motions of stars in the fields, we simply
calculated the individual differences of the proper motions of
stars from two catalogues, and then we studied the distribution of
these differences on the field. These dependencies for the
individual differences of the proper motions (XPM1.0--UCAC3.0) are
shown in Fig. 5.

As seen from the Fig. 5 the individual differences of proper
motions of stars have an unnatural behavior. In our opinion, the
proper motions of stars should not display such an unnatural
behavior within the relatively small field of about 5 by 5 degree.
We can expect the linear dependence or small quadratic
nonlinearity at a pinch. Therefore we believe that this behavior
is non-real and most likely is caused by the systematic positional
errors of catalogues. In order to clear up which catalogue the
majority of these errors belongs to, we constructed the
dependences of the proper motions versus the coordinates for the
XPM-1.0 (Fig. 7) and the UCAC-3.0 (Fig. 6) catalogues separately.

As can be seen in the Fig. 6 the UCAC-3.0 catalogue contains
remarkable systematic errors. An analysis of behavior of proper
motions UCAC-3.0 stars in various fields have shown that in
certain areas of the sky, these stepwise discontinuity can reach a
considerable value to 20--30 mas/yr. Despite the declared accuracy
that UCAC-3.0 catalogue has very small errors an average across
the sky, it appears that in most cases in the fields of the sky
with size of 5 by 5 degree (especially in the northern hemisphere)
the unnatural behavior of proper motions is observed, which
indicates, in our opinion, that the stepwise discontinuity
behavior of proper motions in the catalogue are not excluded.
These errors in some fields may be very significant. This fact is
important, because most modern observations with CCDs are
performed in small-sized fields, where the reference stars can
have unfortunate systematic errors.

To obtain  systematic differences of proper motions and their
dispersions  depending on the magnitude, the range of stellar
magnitudes was divided into the sub-bands with width of 0.05
magnitude. Then, in each of these sub-bands the differences of
proper motions, as well as their dispersions were calculated. The
dependencies of systematic differences of proper motions between
catalogues UCAC-2.0, UCAC-3.0 and XPM-1.0 are shown in Fig. 8 and
Fig. 9 for the northern and southern hemisphere respectively.
Undoubtedly, the systematic differences of proper motion
(UCAC2.0--UCAC3.0) for the northern and southern hemisphere
respectively are the most intriguing feature. The appearance of
the systematic differences between the proper motions of UCAC-2.0
and UCAC-3.0 can be due to the using of early epoch SPM data
($-90^\circ$ to $-10^\circ$ Dec) and  Schmidt plates data from the
SuperCOSMOS project. As may be seen in the figures given, the
standard deviation for northern hemisphere is approximately 8
mas/yr under comparison with UCAC-2.0 and 14 mas/yr under
comparison with UCAC-3.0 in the range from 14 to 16 magnitudes,
where we suppose that no the magnitude equation in the XPM-1.0
catalogue exists. For southern hemisphere the standard deviation
is approximately 16--18 mas/yr under comparison with UCAC-2.0 and
15--16 mas/yr under comparison with UCAC-3.0. Unfortunately, the
use of internal errors of proper motions of the both catalogues
yields a result that is not consistent with the values of standard
deviations of proper motions presented in Fig. 9. Even if we use
the maximum values of the internal errors of proper motions,
stated in catalogues: 8 mas/yr for UCAC-3.0 and 10 mas/yr for the
XPM-1.0, the result does not exceed 13 mas/yr. Thus, a comparison
of the XPM-1.0 and UCAC-2.0, UCAC-3.0 with the aim to determine
the external errors of the proper motion in both catalogues
separately shows that the internal error in one of them or in both
is defined incorrectly.

In order to estimate the external errors of proper motions of
XPM-1.0 catalogue we intended  to use the statistical method of
errors calculation, proposed by Wielen (1995). The method is based
on a comparison of sufficient number of independent proper motions
and positions. However, because the Schmidt plates data from the
project SuperCOSMOS were used for derivation of the UCAC-3.0
proper motions this intention was not feasible.

The discovered systematic difference in proper motions can be
caused by the rotation of UCAC-2.0, 3.0 systems and XPM-1.0 system
each relative to other. However, for the final conclusion the
XPM-1.0 catalogue should be carefully studied and the magnitude
equation and color equation should be securely excluded in whole
range of stellar magnitudes.

\begin{figure*}
\includegraphics[width = 155mm,]{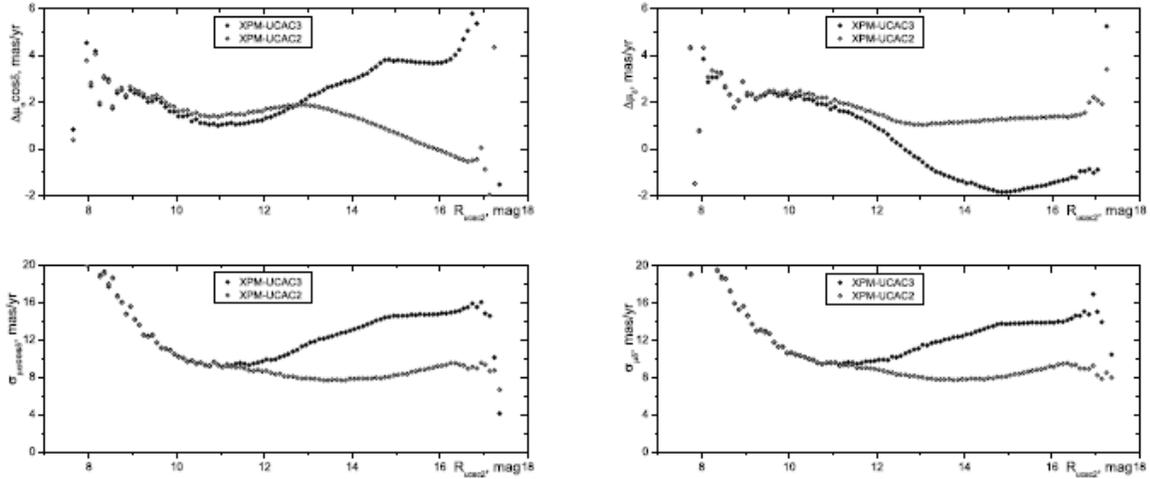}
\caption{The systematic differences of proper motions and their
standard deviations (XPM1.0--UCAC3.0, XPM1.0--UCAC2.0) in the
northern hemisphere as a function of magnitude $R_{UCAC2.0}$}
\end{figure*}

\begin{figure*}
\includegraphics[width = 155mm,]{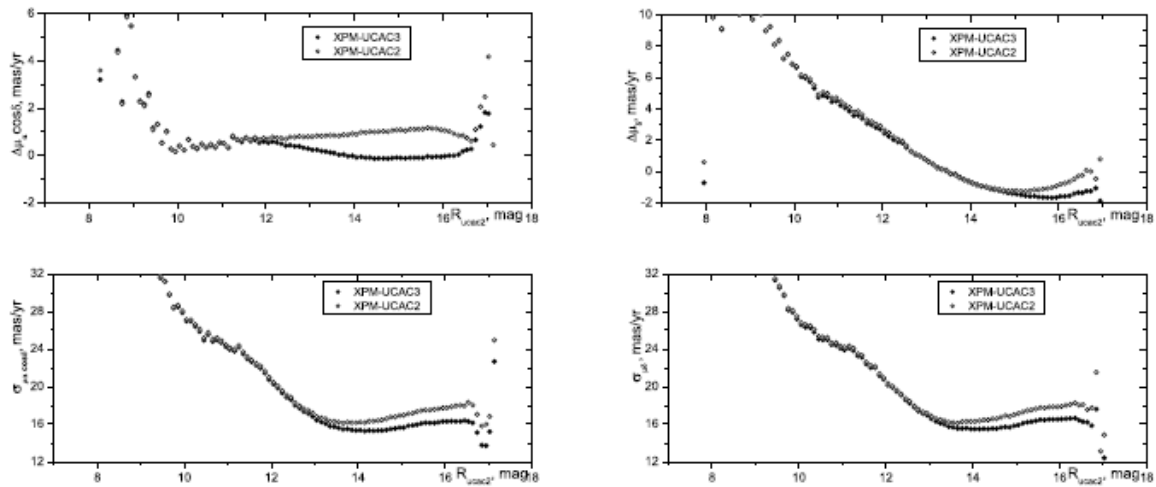}
\caption{The systematic differences of proper motions and their
standard deviations (XPM1.0--UCAC3.0, XPM1.0--UCAC2.0) in the
southern hemisphere as a function of magnitude $R_{UCAC2.0}$}
\end{figure*}

\section{Properties of the XPM-1.0 catalogue}
This version of the XPM catalogue contains the original absolute
proper motions of about 280 million stars. Most of these absolute
proper motions have been determined for the first time. As we
noted earlier, the absolute calibration accuracy for the northern
and southern hemispheres is unequal. This is caused not only by
the lesser mean difference the of epochs for the southern
hemisphere, but due to the different amount of galaxies contained
in these hemispheres as well. The XSC catalogue contains about 1
million galaxies for the northern hemisphere, whereas about 0.5
million galaxies are included for the southern one. This
proportion is retained for the XPM-1.0 catalogue. The XPM-1.0
positions  were calculated for the mean epoch of a concrete object
as the average values of the source 2MASS position  and its
USNO-A2.0 one  reduced to the 2MASS system after applying the
median filter. As the 2MASS positions are tied to the ICRS system,
the XPM-1.0 catalogue contains the formal ICRS positions of all
objects reduced by means of the proper motions to the epoch J2000.
Moreover, it should be noted that as to those objects that occur
twice in the overlapping USNO-A2.0 fields, their positions and
proper motions were obtained by a simple averaging the positions
and proper motions in the intersection. We did not  classify using
the discernibility criterion for stellar or non-stellar objects,
as it was done, for example in the GSC2.3 catalogue (Lasker et
al., 2008). The flag indicating that the extended source was put
into the catalogue was introduced only for the XSC objects. It
seems to us that the number of stars with absolute proper motions
contained in the XPM-1.0 catalogue is the reasonable and
practically coincides with the number of stellar objects
($\approx$210 millions) in the GSC2.3 catalogue which includes
data for about 1 billion objects contained in the Schmidt plates.
The XPM-1.0 catalogue covers the entire sky in the range of
stellar magnitudes $10^{m}<B<22^{m}$ and contrary to the previous
version it does not have any gaps in the zone of the galactic
equator. For each XPM-1.0 object the $J, H, K, B, R$ magnitudes
and their errors are taken from  the corresponding catalogues
containing these quantities.

\section{Conclusions}
The main goal of this work is to provide an independent
realization of the quasi-inertial reference frame which based on
the catalogue of absolute proper motions of 280 millions stars and
can be used for many astronomical studies. As is well known the
zone of avoidance is of great interest for astrophysics and
stellar astronomy. Therefore, for fields from this zone of
avoidance or which contain less than 25 galaxies, we applied a
procedure called by us a quasi absolute calibration. The
parameters of the reduction model were obtained by interpolation
of the values from neighboring of the fields. At this point, we
have done a more thorough identification of objects in the source
catalogues. This allows to decrease the number of false stars and
to improve the quality of the absolute calibration. Besides, we
have made more detailed analysis of the obtained results in order
to investigate of the magnitude equation and  comparison the
proper motions with those contained in the recent catalogues. We
have found a systematic difference between the proper motions in
catalogue XPM-1.0 and UCAC-2.0, UCAC-3.0, which reaches several
mas/yr. The existence of the systematic differences between the
UCAC-2.0 and UCAC-3.0 is the most surprising fact. This fact
hampers to obtaining an objective estimate accuracy by comparing
the catalogues. It is obvious that the internal estimates of
accuracy of proper motions compared catalogues are too low in
either one of them or both catalogues and additional research are
required. Since we will assume further studies of the proper
motions in the bright end of the range of the XPM-1.0 catalogue
stellar magnitudes in order to identify and eliminate the
magnitude equation, we have left wittingly the magnitude equation
in the bright part of the XPM-1.0 catalogue for a while unchanged.
Analysis of the behavior of proper motions of UCAC-3.0 stars in
various fields has shown that in certain areas of the sky have the
stepped discontinuities, reaching 20--30 mas/year. This fact
should be borne in mind because most modern observations with CCDs
are performed in small-sized fields, where the reference stars can
have unfortunate systematic errors.

We have almost completed the preparation of the catalogue XPM for
release and hope that the final version of the catalogue XPM will
be available via CDS in Strasbourg during 2010. Currently, for
access to a intermediate version of XPM-1.0, you may contact with
Fedorov P.N. or Akhmetov V.S. by  e-mail: {\tt
pnf@astron.kharkov.ua} or {\tt akhmetov@astron.kharkov.ua}.

\section*{ACKNOWLEDGMENTS}

This study was supported by the Fundamental Researches State Fund
of Ukraine (project No. FRSF--28/238) and the Russian Foundation
for Basic Research (project No. 08--02--00400 and No.
09--02--90443--Ukr\_f), and in part by the ``Origin and Evolution
of Stars and Galaxies'' --- ``Program of the Presidium of the
Russian Academy of Sciences and the Program for State Support of
Leading Scientific Schools of Russia'' (NSh--3645.2010.2).

{

}

\end{document}